\documentclass[11pt]{article}

\usepackage[T1]{fontenc}
\usepackage[sc]{mathpazo}
\usepackage{amsmath}
\usepackage{amssymb}
\usepackage{enumerate}
\usepackage{amsthm}

\usepackage{amsfonts,mathrsfs}
\usepackage[style]{fncychap}
\usepackage{graphicx} 
\usepackage{geometry} 
\usepackage{eepic}
\usepackage{ifthen}
\newboolean{ElectronicVersion}
\setboolean{ElectronicVersion}{true} 

\usepackage{hyperref}
\hypersetup{pdfpagemode=UseNone}

\def\be{\begin{equation}}
\def\ee{\end{equation}}
\def\bea{\begin{eqnarray*}}
\def\eea{\end{eqnarray*}}

\newenvironment{mylist}[1]{\begin{list}{}{
    \setlength{\leftmargin}{#1}
    \setlength{\rightmargin}{0mm}
    \setlength{\labelsep}{2mm}
    \setlength{\labelwidth}{8mm}
    \setlength{\itemsep}{0mm}}}
    {\end{list}}

\newcommand{\bra}[1]{\langle#1|}

\newcommand{\ket}[1]{|#1\rangle}

\theoremstyle{definition}

\numberwithin{equation}{section}

\newcounter{questionnumber}

\begin{document}

\title{\Large Field extension of real values of physical observables in classical theory \\can help attain quantum results}


\author{Hai Wang\footnote{
3090101669@zju.edu.cn}\\
  {\small \it School of Mathematics Sciences, Zhejiang University, Hangzhou 310027, PR~China}\\
Asutosh Kumar\footnote{usashrawan@gmail.com}\\
  {\small \it Harish-Chandra Research Institute, HBNI, Chhatnag Road, Jhunsi, Allahabad 211019, India}\\
  {\small \it P. G. Department of Physics, Gaya College, Magadh University, Rampur, Gaya 823001, India}\\Junde Wu\footnote{wjd@zju.edu.cn}\\{\small \it School of Mathematics Sciences, Zhejiang University, Hangzhou 310027, PR~China}\\Minhyung Cho\footnote{Correspongding author: mignon@kumoh.ac.kr}\\{\small \it Department of Applied Mathematics, Kumoh National Institute of Technology, Kyungbuk,
730-701, Korea}}

\date{}
\maketitle \mbox{}\hrule\mbox\\
\begin{abstract}
 Physical quantities are assumed to take real values, which stems from the fact that an usual measuring instrument that measures a physical observable always yields a real number. Here we consider the question of what will happen if physical
observables are allowed to take complex values. In this paper, we show that by allowing observables in the Bell
inequality to take complex values, a classical physical theory can
actually get the same upper bound of the Bell expression as quantum
theory. Also, by extending the real field to the quaternionic field,
we can puzzle out the GHZ problem using local hidden variable model.
Furthermore, we try to build a new type of hidden-variable theory of
a single qubit based on the result.
\end{abstract}
\mbox{}\hrule\mbox\\

\section{Introduction}
We have learnt from experience that measured variables yield real values. Therefore physical quantities are assumed to take real values because a typical measuring apparatus that measures a physical observable always yields a real number. In quantum theory, a physical observable is represented by a hermitian operator as it admits a spectral decomposition and has real eigenvalues. However, the real value of all physical quantities is an assumption only. Moreover, the real value of eigenvalues of some operator does not mean that it is necessarily hermitian. There are examples of non-hermitian operators which possess real eigenvalues under some
symmetry conditions, namely the ${\cal PT}$ symmetry \cite{pt-sym1,pt-sym2,pt-sym3}. We ask the following question: Can the outcomes of a physical observable be complex? It may be the limitation of a measuring apparatus that it measures only real values, or we have not invented a way out to naturally measure complex values. In general, given a non-hermitian operator, its average value in a quantum state is a complex number.
In quantum theory, the concept of {\em weak measurement} was introduced by Aharonov {\em et al.} \cite{weak-meas1,weak-meas2,weak-meas3} to study the properties of a quantum system in pre- and postselected ensembles. In that formalism, the measurement of an observable yields a weak value of the observable with unusual properties. In general, the weak value is complex, and can take values outside the eigenspectrum of the observable.
In recent years, weak values have found several applications \cite{weak-app1,weak-app2,weak-app3,weak-app4,weak-app5,weak-app6,weak-app7,weak-app8,weak-app9,weak-app10,weak-app11,weak-app12,weak-app13,weak-app14}.
One can see \cite{weak-rev} for a recent review on weak measurements.

However, as a mathematical structure, unlike the real field, the field of complex numbers does not admit an {\em order relation}. Therefore, it is more difficult to compare ``observables'' having complex numbers as eigenvalues.
One might, however, argue that complex numbers can be represented as pairs of real numbers, and measure them separately. More generally, one can measure quantities which are expressible as an array of real numbers. This is alright in classical mechanics.
However, one must remember, as already noted by Dirac \cite{dirac}, that measuring an ``observable'' with complex eigenvalues by somehow measuring separately its real and pure imaginary parts could lead to problems in the quantum formalism, as two observations may ``interfere'' with one another.

In spite of having complex eigenvalues, non-hermitian operators have
found several applications
\cite{nonhermitian-app1,nonhermitian-app2,nonhermitian-app3,nonhermitian-app4,nonhermitian-app5,nonhermitian-app6,nonhermitian-app7,nonhermitian-app8,nonhermitian-app9,nonhermitian-app10,nonhermitian-meas}
in studying open quantum systems in nuclear physics
\cite{nonhermitian-app1} and quantum optics
\cite{nonhermitian-app2}, to name a few.
Regardless of how a complex value could be measured or what would be
its interpretation, here we consider the question of what would
happen if physical observables are allowed to take complex values.
Bender and Boettcher \cite{pt-sym1} showed that by allowing physical
quantities to take complex values, classical particles can behave
much like particles in quantum theory. In this paper, we show that
by allowing observables in the Bell inequality to take complex
values, a classical physical theory can actually get the same upper
bound of the Bell expression, as in quantum theory. Moreover, using
the local hidden variable model, and by extending the real field to
the quaternionic field, we visit the GHZ paradox and show that there
exists a non-empty set of 9-tuples of quaternions which correspond
to negative eigenvalue of the GHZ argument.

\section{Bell inequality and the corresponding upper bound}
\label{sec-bell} In this section, we show that if the observables are allowed to take complex
values, then the quantum upper bound on the
Bell-Clauser-Horne-Shimony-Holt expression \cite{bell-ineq} can be obtained by using the traditional hidden variable method, too.

\subsection{Bell inequality}
Bell inequality is a mathematical inequality involving certain averages of correlations of measurements, derived using the assumptions of locality and realism.
In classical physics, for a system, all observables should have
exact values before measurements and different systems cannot affect
each other instantly any more once they are separated. This is what
locality and realism mean and the core belief of classical physics.
Bell showed that quantum mechanics violates this inequality. That
is, quantum mechanics cannot be both local and realistic. The
setting is as follows. There are two observers, Alice (A) and Bob
(B). Each of them has two experimental settings: A(a) and A(a') for
Alice, and B(b) and B(b') for Bob. And all these observables are
dichotomic, that is, taking values from $\{1£¬ -1\}$.

In classical physics, for a system, physical quantities may have many different values and it is the system's state that decides which value these physical quantities take. And thinking classically, if values of some quantities show randomness for a system, then to explain possible randomness
in experiments, there should be some unknown variable $\lambda \in
\Lambda$, which helps us to give a complete description for a sytem's state and determines the value of each quantity, and which may
in principle remain hidden from observation with currently available
measuring technology. Such variable is usually referred to as hidden
variables. Each run of an experiment corresponds to a realization of
the random variable $\lambda$ with a probability distribution
$\rho(\lambda)$. The correlation function $E(a,b)$ of the experiment
which involves the measurements of $A(a)$ and $B(b)$ at Alice and
Bob's laboratories is defined by
\begin{equation}
\label{eq-corr} E(a,b)=\int d\lambda
\rho(\lambda)A(a,\lambda)B(b,\lambda).
\end{equation}
Here, $A(a,\lambda)$ denotes the value of the observable $A(a)$ corresponding to $\lambda$. Strictly, the hidden variable $\lambda$ in $A(a, \lambda)$ decides which value this quantity actually takes. The famous Bell inequality states that
\begin{equation}
\label{eq-bell}
|E(a,b)-E(a,b')|+|E(a',b)+E(a',b')|\leq 2.
\end{equation}
This inequality is valid in any physical theory that is local and
realistic, where the involved physical observables take values from $\{+1, -1\}$. We
refer to the expression on the left-hand-side of (\ref{eq-bell}) as
the Bell expression. In any theory, if the Bell expression exceeds
the value 2, then the theory cannot be both local and realistic. Quantum
mechanics happens to be such a theory.

\subsection{Real values to complex values}
It follows from above that in a typical Bell experiment,
classical physics permits us to get the maximum value 2, while
quantum physics can make us have the maximal value 2$\sqrt{2}$,
which is known as Tsirelson's bound \cite{tsirelson}. Now, we allow the observables to take complex values, that is, the observable $A(a)$ take complex values and the
hidden variable $\lambda$ is employed as follows:
\begin{align}
\label{eq-complex}
A(a,\lambda)&=(-1)^{f_{1}(\lambda)}e^{i\theta_{1}},\,\,\,\,\,\, A(a',\lambda)=(-1)^{f_{3}(\lambda)}e^{i\theta_{3}}\\
\nonumber
B(b,\lambda)&=(-1)^{f_{2}(\lambda)}e^{i\theta_{2}},\,\,\,\,\,\,
B(b',\lambda)=(-1)^{f_{4}(\lambda)}e^{i\theta_{4}},
\end{align}
where functions $f_{i}(\lambda)=0$ or $1$, $i=1,2,3,4$.

Following the original derivation of the Bell inequality, we have:
$$|E(a,b)-E(a,b')|$$
$$=|\int d\lambda\rho(\lambda)(-1)^{f_{1}(\lambda)}e^{i\theta_{1}}(-1)^{f_{2}(\lambda)}e^{i\theta_{2}}$$
$$-\int d\lambda\rho(\lambda)(-1)^{f_{1}(\lambda)}e^{i\theta_{1}}(-1)^{f_{4}(\lambda)}e^{i\theta_{4}}|$$
$$\leq \int d\lambda\rho(\lambda)|(-1)^{f_{2}(\lambda)}e^{i\theta_{2}}-(-1)^{f_{4}(\lambda)}e^{i\theta_{4}}|,$$
as $|(-1)^{f_{1}(\lambda)}e^{i\theta_{1}}|=1$.
And similarly,
$$|E(a',b)+E(a',b')|$$
$$\leq \int d\lambda\rho(\lambda)|(-1)^{f_{2}(\lambda)}e^{i\theta_{2}}+(-1)^{f_{4}(\lambda)}e^{i\theta_{4}}|.$$
Hence, $$|E(a,b)-E(a,b')|+|E(a',b)+E(a',b')|$$
\begin{equation}
\label{eq-bound}
\leq |e^{i\theta_{2}}+e^{i\theta_{4}}|+|e^{i\theta_{2}}-e^{i\theta_{4}}|\leq 2\sqrt2.
\end{equation}

It is amazing that the upper bound $2\sqrt2$ coincides with the
upper bound of the Bell-CHSH inequality in quantum theory. Thus, by
allowing observables in the Bell experiment scenario to take complex
values, a classical physical theory can actually get the same upper
bound of the Bell expression as in quantum theory, where by
classical physical theory we mean a theory that is both local and
realistic.

On the other hand, the upper bound is achievable classically, that
is, we can find some appropriate observables and a probability distribution $\rho(\lambda)$ of the hidden variable $\lambda$ which makes the above inequality become an equality. If we choose $\theta_{2}=0, \theta_{4}=\pi/2, \theta_{1}=7\pi/4,
\theta_{3}=\pi/4$, and $\rho(\lambda)$ is supported on those
$\lambda$ which satisfy $f_{2}=f_{4}=0$ or $f_{2}=f_{4}=1$, then the
upper bound $2\sqrt2$ can actually be achieved.


\section{Hidden-variable theory of a qubit}
As we have seen in section \ref{sec-bell}, when we allow physical quantities to take complex values, classical physics can attain quantum physical limit. Furthermore, when this complex-valued classical physics (we call complex-valued classical physics ``pseudo-classical'' physics) gets the maximum in the Bell inequality, the observables involved can share the same relation with the quantum case (just what the above example says). In this regard, we ask {\em is it possible to iterate the quantum theory in one-qubit case using this pseudo-classical approach?} Below we propose a new type of hidden-variable theory of a single qubit via complexification of physical quantities.
\subsection{Observables}
The critical question is how to convert the hermitian operators on the Hilbert space $\mathbb{C}^2$ to functions of hidden variables. Because all hermitian operators can be expressed as a real linear combination of the identity operator and spin operators, so the most important nontrivial observables in the one-qubit case are spins. Following the above complexification of quantities in Bell-CHSH expression, if the spins are in the x-y plane, we can make correspondence as follows:
$$\sigma_{\overrightarrow{\alpha}}\Leftrightarrow f_{\overrightarrow{\alpha}}(\lambda)=(-1)^{g_{\overrightarrow{\alpha}}(\lambda)} e^{i\alpha},$$
where $\alpha$ is the angle of ${\overrightarrow{\alpha}}$ with the $x$ axis and the function $g_{\overrightarrow{\alpha}}(\lambda)$ takes value from $\{0, 1\}$.

The change of $\pm 1$ to antipodal points on the circle is natural, more or less. However, the difficulty arises, because spin-operators of the same plane have exhausted all antipodal points on the complex plane and yet all spin-operators of one qubit are apparently not on the same plane. How can we define functions for spin-operators on the different planes?
Meanwhile, the standard quantum theory tells us that Pauli spin-operators $\sigma_{x}, \sigma_{y}, \sigma_{z}$ have the same important place when we compare them with each other. So, this equivalence should also be embodied in the way we use functions to replace them. This forces us to adopt a further step in complexification. That is, we should introduce quaternions into the pseudo-classical physics:
$$\sigma_{x}\Leftrightarrow f_{x}(\lambda)=(-1)^{g_{x}(\lambda)} i, \sigma_{y}\Leftrightarrow f_{y}(\lambda)=(-1)^{g_{y}(\lambda)} j, \sigma_{z}\Leftrightarrow f_{z}(\lambda)=(-1)^{g_{z}(\lambda)} k,$$where functions $g_{x}(\lambda)$, $g_{y}(\lambda)$ and $g_{z}(\lambda)$ take values from $\{0, 1\}$. Thereof, quaternions are numbers of the following form:$$a+bi+cj+dk,$$with a,b,c,d being real. And $i, j, k$ are three different imaginary units, satisfying the following relationship$$i\cdot j=k,\ j\cdot k=i,\ k\cdot i=j, i^{2}=j^{2}=k^{2}=-1.$$
Then for an arbitrary spin-operator $\sigma_{\overrightarrow{r}}= \overrightarrow{r}.\overrightarrow{\sigma}=\sum_{i=1}^3 r_{i}\sigma_{i}$ with an unit vector $\overrightarrow{r}=(r_{x}, r_{y}, r_{z})^{T} \in \mathbb{R}^3$, what we do is the following:
$$\sigma_{\overrightarrow{r}} \Leftrightarrow f_{\overrightarrow{r}}(\lambda)=(-1)^{g_{\overrightarrow{r}}(\lambda)}(r_{x}i+r_{y}j+r_{z}k),$$where $g_{\overrightarrow{r}}(\lambda)$ takes value from $\{0, 1\}$.

Apparently, the identity operator appearing in the formal quantum theory should be replaced by the function $f(\lambda)\equiv 1$. Now for any hermitian operator on $\mathbb{C}^{2}$, we can use a quaternion-valued function to replace it using the aboved method.

Since physical quantity can take quaternions in this pseudo-classical model, then why do measurements yield only real values? Consider the measurement of spin in the x direction, for example. We can imagine the measurement's action as a fictitious rigid pole. Considering the sphere of unit ball in the i-j-k space, when we say that we are measuring the spin on the x direction, it means that the fictitious pole is along the x direction. Only when the fictitious pole is on the same direction as the spin is, it can make the value of the corresponding function real. After all, it is an essential requirement in the standard quantum mechanics that the result of a measurement be real.

Once we accept this model, then things in quantum mechanics may be somewhat natural. We all know that spins on different directions cannot be measured at the same time. Standard quantum mechanics tell us that it is because of the noncommutativity of those corresponding operators that we cannot measure them simultaneously. However, in our model, the explanation is quite apparent. It is that the fictitious rigid pole of the measurement apparatus cannot be placed in different angles at the same time.
\subsection{States}
\label{sec-states}
When we talk about states in a hidden-variable theory, this means that we should give both the range of hidden variables and the probability distribution corresponding to each state in the standard theory. That is exactly what we do next.

For a one-qubit state, in the standard theory, every quantum state can be expressed as$$\rho=\frac{1}{2}(I+\overrightarrow{r}\cdot \overrightarrow{\sigma}).$$So if we know expectations of $\sigma_{x}$, $\sigma_{y}$ and $\sigma_{z}$, then we can reconstruct this state and predict the result about any other observable. So, considering the equal importance of these three spin-operators, the measurable set of hidden-variable $\lambda$ should be devided into eight disjoint sets of the same measure, as shown in Table \ref{table:8sets}. For example, set 1 consists of those $\lambda$, which satisfy
$$f_{x}(\lambda)=i, f_{y}(\lambda)=j, f_{z}(\lambda)=k,$$
and set 8 is consisted of those $\lambda$, which satisfy
$$f_{x}(\lambda)=-i, f_{y}(\lambda)=-j, f_{z}(\lambda)=-k.$$
So, in our hidden variable model, we can assume that the range of $\lambda$ is the set
$$\lambda \in \{ 1,2,3,4,5,6,7,8\}.$$

Once this is done, we can give the probability distribution of each state of a qubit. Firstly, consider some simple instances.

$$\ket{+_{x}}\bra{+_{x}}\Rightarrow P_{+_{x}}(\lambda)=\frac{1}{4}, \forall \lambda=1,2,3,4;~~~ \ket{-_{x}}\bra{-_{x}}\Rightarrow P_{-_{x}}(\lambda)=\frac{1}{4}, \forall \lambda=5,6,7,8$$
$$\ket{+_{y}}\bra{+_{y}}\Rightarrow P_{+_{y}}(\lambda)=\frac{1}{4}, \forall \lambda=1,2,5,6;~~~ \ket{-_{y}}\bra{-{y}}\Rightarrow P_{-_{y}}(\lambda)=\frac{1}{4}, \forall \lambda=3,4,7,8$$
$$\ket{+_{z}}\bra{+_{z}}\Rightarrow P_{+_{z}}(\lambda)=\frac{1}{4}, \forall \lambda=1,3,5,7;~~~ \ket{-_{z}}\bra{-_{z}}\Rightarrow P_{-_{z}}(\lambda)=\frac{1}{4}, \forall \lambda=2,4,6,8$$

The reason we design this correspondence is that for spins on the x, y, z directions, if the state is an eigenvalue of one, then we could know nothing about the other two. Take the probability distribution $P_{+_{x}}$ as an example. In the above, we use it to correspond to the quantum state $\ket{+_{x}}$, which is the eigenstate of $\sigma_{x}$ corresponding to the eigenvalue $+1$. For state $\ket{+_{x}}$, its probability distribution corresponding to observables $\sigma_{y}$ and $\sigma_{z}$ are all uniform distributions, meaning it can tell us nothing about $\sigma_{y}$ and $\sigma_{z}$. On the other hand, the probability distribution $P_{+_{x}}$ is supported on the subset $\{1,2,3,4\}$ of the hidden variables. In terms of Table 1, hidden variables in this subset always make the function $f_{x}$ get the value $i$. And furthermore, taking the probability distribution $P_{+_{x}}$ into consideration, its marginal distributions about functions $f_{y}$ and $f_{z}$ are uniform distributions as well. Based on these results, we make the above correspondence.

Then, for an arbitrary state, can we give its probability distribution? For a qubit, we know that an arbitrary state can be expressed as:
$$\rho=\frac{1}{2}(I+\overrightarrow{r}\cdot \overrightarrow{\sigma}),$$where $\overrightarrow{r}=(r_{x}, r_{y}, r_{z})\in R^{3}$ satisfies $\|\overrightarrow{r}\|\leq 1$.

At a first glance, it's natural to guess that the probability distribution of an arbitrary state $\rho$ is (mapping $\sigma_{x}=\ket{+_{x}}\bra{+_{x}}-\ket{-_{x}}\bra{-_{x}}\rightarrow P_{+_{x}}-P_{-_{x}}$):

$$P_{\rho}=\frac{1}{2}[1+r_{x} (P_{+_{x}}-P_{-_{x}})+r_{y} (P_{+_{y}}-P_{-_{y}})+r_{z} (P_{+_{z}}-P_{-_{z}})].$$
But this is not normalised. The normalized probability distribution for arbitrary $\rho$ is \cite{note}:
$$P_{\rho}=\frac{1}{2}[1/4+r_{x} (P_{+_{x}}-P_{-_{x}})+r_{y} (P_{+_{y}}-P_{-_{y}})+r_{z} (P_{+_{z}}-P_{-_{z}})].$$

With this probability distribution, using our method, it's quite easy to check that this kind of probability distribution allows us to get the same results about the three observables $\sigma_{x}$, $\sigma_{y}$ and $\sigma_{z}$ as we get in the standard way. Since, for a qubit state, once we know its probability distributions about the three spins, then we can reconstruct the state, or in other words, we can predict its probability distribution about any observable. Hence, it's reasonable to expect that our probability distributions of hidden variables can make the same prediction as those made by the standard quantum theory, as these two different methods share a complete agreement on the three critical observables. Discussion about this will be put in section \ref{sec-discussions}.

\begin{center}
\begin{table}[htb]
\begin{tabular}{|c|c|}\hline
set no. $(\lambda)$ & $f_{x}(\lambda), f_{y}(\lambda), f_{z}(\lambda)$ \\ \hline
1 & i, j, k \\ \hline
2 & i, j, -k \\ \hline
3 & i ,-j, k \\ \hline
4 & i, -j, -k \\ \hline
5 & -i, j, k \\ \hline
6 & -i, j, -k \\ \hline
7 & -i, -j, k \\ \hline
8 & -i, -j, -k \\ \hline
\end{tabular}
\caption{
Any observable $A=a I+r_x \sigma_x+r_y \sigma_y+r_z \sigma_z$, of a qubit is a hermitian operator where $a \in \mathbb{R}$ and $(r_x,r_y,r_z)\in \mathbb{R}^{3}$.
In the view of hidden-variable theory, it seems that once the hidden variable $\lambda$ has determined the values of $\sigma_{x}$, $\sigma_{y}$ and $\sigma_{z}$, the value of any other observable is determined. This inspires us to divide the range of $\lambda$ into eight disjoint sets, depending on the values of $\sigma_{x}\Rightarrow f_{x}(\lambda)=\pm i, \sigma_{y}\Rightarrow f_{y}(\lambda)=\pm j, \sigma_{z}\Rightarrow f_{z}(\lambda)=\pm k$. Of course, these eight sets should have the same ``area''. This is why we can assume that the hidden variable $\lambda$ is chosen in the set $\{ 1,2,3,4,5,6,7,8\}$.
}
\label{table:8sets}
\end{table}
\end{center}

Until now, it seems that this method is quite good. But there is something unsatisfactory. If $\overrightarrow{r}=(\frac{1}{\sqrt{3}}, \frac{1}{\sqrt{3}}, \frac{1}{\sqrt{3}})^{T}$, then the corresponding probability distribution is somewhat annoying. Based on easy computation,
$$P_{\overrightarrow{r}}(\lambda=8)=\frac{1}{8}(1-\sqrt{3})< 0.$$
Can we find reasonable explanation for this negative probability?
We interpret this negative item as the augmentation of its ``retroaction''. What does ``retroaction" mean? From Table \ref{table:8sets}, because set 8 corresponds to those $\lambda$, which satisfy
$$\lambda\in \{\lambda\mid f_{x}(\lambda)=-i, f_{y}(\lambda)=-j, f_{z}(\lambda)=-k \}.$$
So set 8's "retroaction" is set 1.

In concrete terms, for an arbitrary quantum state $\rho$, its probability distribution must observe the following constraints:
$$P_{\rho}(m)+P_{\rho}(9-m)=\frac{1}{4}, \forall m=1,2,3,4.$$
Following these constraints, the word "retroaction" may be appropriate. In this way, the negative probability causes no trouble.

\section{The GHZ argument}
\label{sec-ghz} In this section, by using our hidden variable model, we attempt to view the GHZ problem in another new way.

\subsection{Original GHZ argument}
The GHZ argument \cite{ghz-argument} states that the quantum state,
$\ket{\Psi}=\frac{1}{\sqrt{2}}(\ket{000}-\ket{111})$, satisfies
\begin{equation}
\label{eq-ghz-cond}
O_{i}\ket{\Psi}=+\ket{\Psi},
\end{equation}
where $i\in\{1,2,3\}$, $\{\sigma_{x},\sigma_{y},\sigma_{z}\}$ are
the Pauli operators, $\{\sigma^i_{x},\sigma^i_{y},\sigma^i_{z}\}$
are the Pauli operators $\{\sigma_{x},\sigma_{y},\sigma_{z}\}$ on
the $i$-th subsystem, $O_{1}=\sigma^{1}_{x}\otimes
\sigma^{2}_{y}\otimes \sigma^{3}_{y} \equiv
\sigma^{1}_{x}\sigma^{2}_{y}\sigma^{3}_{y},\
O_{2}=\sigma^{1}_{y}\sigma^{2}_{x}\sigma^{3}_{y},\
O_{3}=\sigma^{1}_{y}\sigma^{2}_{y}\sigma^{3}_{x}.$

In the traditional LHV model, each observable is represented by its
possible values $S^{m}_{\alpha}$, not by its operator. then the
conditions (\ref{eq-ghz-cond}) translate
to:$$S^{1}_{x}S^{2}_{y}S^{3}_{y}=+1=S^{1}_{y}S^{2}_{x}S^{3}_{y}=S^{1}_{y}S^{2}_{y}S^{3}_{x},$$
where $S^{m}_{\alpha}\in \{1,-1\},\ \ m\in \{1,2,3\},\ \ \alpha\in
\{x,y\}$. Thus, we have
\begin{equation}
\label{eq-ghz}
S^{1}_{x}S^{2}_{x}S^{3}_{x}=(S^{1}_{x}S^{2}_{y}S^{3}_{y})(S^{1}_{y}S^{2}_{x}S^{3}_{y})(S^{1}_{y}S^{2}_{y}S^{3}_{x})=+1.
\end{equation}

On the other hand, however, in quantum theory, we have
$\sigma^{1}_{x}\sigma^{2}_{x}\sigma^{3}_{x}\ket{\Psi}=-\ket{\Psi}$,
which is in contradiction with Eq. (\ref{eq-ghz}). This shows that
there is a conflict between quantum theory and classical theory
which admits the local realism.

\subsection{Another way to the GHZ problem}

In the classical case, we have $$S^{m}_{n}\in \{1,-1\},\ \ m\in \{1,2,3\},\ \ n\in \{x,y\}.$$
So in the classical reasoning, the condition $S^{1}_{x}S^{2}_{y}S^{3}_{y}=+1$, has four possibilities:
$$S^{1}_{x}=+1,\ S^{2}_{y}=+1, \ S^{3}_{y}=+1\ \ \ (1)
$$$$S^{1}_{x}=+1,\ S^{2}_{y}=-1, \ S^{3}_{y}=-1\ \ \ (2)$$
$$S^{1}_{x}=-1,\ S^{2}_{y}=+1, \ S^{3}_{y}=-1\ \ \ (3)$$
$$S^{1}_{x}=-1,\ S^{2}_{y}=-1, \ S^{3}_{y}=+1\ \ \ (4)$$


Firstly, we consider the possibility (1). Using Table 1, it just means that the hidden variable $\lambda$ is located in the following set:$$\Lambda_{1}=\{1,2,3,4\}\times \{1,2,5,6\}\times \{1,2,5,6\}.$$

In the same way, the possibilities (2)-(4) yield the following sets:
$$\Lambda_{2}=\{1,2,3,4\}\times \{3,4,7,8\}\times \{3,4,7,8\},$$
$$\Lambda_{3}=\{5,6,7,8\}\times \{1,2,5,6\}\times \{3,4,7,8\},$$
$$\Lambda_{4}=\{5,6,7,8\}\times \{3,4,7,8\}\times \{1,2,5,6\}.$$
So the condition $S^{1}_{x}S^{2}_{y}S^{3}_{y}=+1$ corresponds to the union of these four sets, $\bigcup_{i=1}^{4}\Lambda_{i}\equiv \Lambda $.

Similarly, the conditions $S^{1}_{y}S^{2}_{x}S^{3}_{y}=+1$ and $S^{1}_{y}S^{2}_{y}S^{3}_{x}=+1$,
separately correspond to sets $\bigcup_{i=1}^{4}\Pi_{i} \equiv \Pi $ and $\bigcup_{i=1}^{4}\Omega_{i} \equiv \Omega $ respectively. Now we take intersection of the three sets $\Lambda,\ \Pi,\ \Omega$. If we get an empty set, then our method will not solve the GHZ puzzle. If we get some non-empty set, then we have to check whether hidden variables $\lambda$ corresponding to that set can satisfy $S^{1}_{x}S^{2}_{x}S^{3}_{x}=-1$.

Interestingly, we get a non-empty set, which is the union of the following sets:
$$\{1,2\}\times\{1,2\}\times\{1,2\},$$
$$\{1,2\}\times\{7,8\}\times\{7,8\},$$
$$\{7,8\}\times\{1,2\}\times\{7,8\},$$
$$\{7,8\}\times\{7,8\}\times\{1,2\}.$$
For these sets (or, consequently $\lambda$), we can build probability distributions supported on them(in other words, we can build a state since for traditional local hidden variable models, a state is equivalent to a probability distribution on hidden variables). However, note that for these sets, considering $S^{1}_{x},\ S^{2}_{x}$ and $S^{3}_{x}$, we can always get:$$S^{1}_{x}S^{2}_{x}S^{3}_{x}=-i.$$ Compare this result with the quantum case $$\sigma^{1}_{x}\sigma^{2}_{x}\sigma^{3}_{x}\ket{\Psi}=-\ket{\Psi}.$$

Thus, if we extend the real field to the quaternionic field, then even in the LHV model there can be correlations as strong as in quantum mechanics, and can puzzle out the GHZ problem.
\section{Discussions}
\label{sec-discussions}
\subsection{Imperfection}

Let us go back to the question left in section \ref{sec-states}. For our hidden-variable model, if we want our constructed probability distribution $P_{\rho}$ for a given $\rho$ to give the same prediction about any observable as those got in the orthodox way, then it means that: {\em For any hidden-variable $\lambda\in \{1,2,...,8\}$ and any dichotomic valued function $f_{A}(\lambda)$, we can find an appropriate way to decide which value this function should take for each $\lambda$, where the subscript A is an arbitrary hermitian operator on $\mathbb{C}^{2}$ showing that this function is the one corresponding to it in our hidden-variable model}.

However, it's a pity that this appropriate method have not been found.

\subsection{Evolution}
In the standard quantum theory, unitary operators play a vital role in the description of states' evolution. Then, can we also describe states' evolution in our model? From the view of hidden-variable model, every quantum state is a statistical mixture of delicate states, which are indicated by hidden variables $\lambda$. So if we want to simulate unitary evolutions in a hidden variable model, then it should be some kind of mixture of transformations among these hidden states, because in the hidden variable view, the unitary evolution of quantum states, is a global behaviour of mixtures of hidden states. Then we have to decide the transformations among these hidden variables. In our model, for one qubit, we introduce hidden variable $\lambda \in \{1,2,...,8\}$. So there are $8!=40320$ different ways of transformations among these hidden variables. This indicates that we can use elements of the permutation group $\textbf{S}_{8}$ to represent transformations among these hidden variables. As quantum states are statistical mixtures of hidden variables, it's reasonable to assume that evolutions of quantum states are also statistical mixtures of transformations among hidden variables. That is, in our hidden variable model, unitary evolutions can be imitated by convex combinations of elements of the permutation group $\textbf{S}_{8}$. Furthermore, if we put these hidden variables in the order induced by their number and then use matrix to represent elements of the permutation group $\textbf{S}_{8}$, then Birkhoff-von Neumann theorem \cite{watrous} tells us that the set of all convex combinations of elements of the permutation group are just the set of all doubly stochastic matrix of the same order. So in our model, unitary operators may be represented by doubly stochasitc matrix. Following is one example.

In our model, we know that the state $\ket{+_{x}}\bra{+_{x}}$ corresponds to the probability distribution $P_{+_{x}}(\lambda)=\frac{1}{4}, \forall 1\leq \lambda \leq 4$, and the state $\ket{-_{x}}\bra{-_{x}}$ corresponds to the probability distribution $P_{-_{x}}(\lambda)=\frac{1}{4}, \forall 5\leq \lambda \leq 8$. Then we can describe the change from $\ket{+_{x}}\bra{+_{x}}$ to $\ket{-_{x}}\bra{-_{x}}$ as follows, using our language:
$$P_{-_{x}}(\lambda)=P_{+_{x}}(\textbf{s}\lambda), \forall 1\leq \lambda \leq 8.$$
The $\textbf{s}$ mentioned above is an element of the group $\textbf{S}_{8}$, which makes the exchange between $i$ and $i+4$ for $\forall 1\leq \lambda\leq 4.$ This is the way the group $\textbf{S}_{8}$ enters into our description of evolution. Then because of the infinite ways of evolution in quantum theory, convex combinations of elements in $\textbf{S}_{8}$ come into being, in terms of probability distributions' normalization and the finiteness of elements in $\textbf{S}_{8}$.

\section{Conclusion}

It seems that quantum theory is a complexification of the classical theory. It may be the strong mutual interaction between physical systems and the measurement apparatus that causes the information stored in the angles lost, which makes the classical theory classical.

In this paper, we investigated to what degree classical theory can behave like quantum theory, when we allow physical quantities to take complex (quaternion) values. Surprisingly enough, we found that in the case of the Bell-CHSH experiment, classical theory yields the same upper bound as the quantum theory. Moreover, by employing LHV model and by extending the real field to the quaternionic field, we can puzzle out the GHZ problem. These results motivated us to build the ``crazy'' hidden-variable theory of a single qubit. We believe that this ``complexification'' method will make useful and interesting contribution in future research.

\section{acknowledgements}
AK is indebted to Ujjwal Sen for illuminating discussions and reading the manuscript, and acknowledges research fellowship from the Department of Atomic Energy, Government of India. The project is supported by Research Fund, Kumoh National Institute of Technology, Korea.


\begin{thebibliography}{99}

\bibitem{pt-sym1} C. M. Bender and S. Boettcher, Phys. Rev. Lett. {\bf 80}, 5243 (1998).
\bibitem{pt-sym2} C. M. Bender, D. C. Brody, and H. F. Jones, Phys. Rev. Lett. {\bf 89}, 270401 (2002).
\bibitem{pt-sym3} D. C. Brody and E.-M. Graefe, Phys. Rev. Lett. {\bf 109}, 230405 (2012).
\bibitem{weak-meas1} Y. Aharonov, D. Z. Albert, and L. Vaidman, Phys. Rev. Lett. {\bf 60}, 1351 (1988).
\bibitem{weak-meas2} Y. Aharonov and A. Botero, Phys. Rev. A {\bf 72}, 052111 (2005).
\bibitem{weak-meas3} Y. Aharonov and L. Vaidman, in {\em Time in Quantum Mechanics}, Lecture Notes in Physics, Vol. {\bf 734} (2008) pp. 399-447.
\bibitem{weak-app1} E. Sjoqvist, Physics Letters A {\bf 359}, 187 (2006).
\bibitem{weak-app2} Y. Aharonov, S. Popescu, and J. Tollaksen, Physics Today {\bf 63}, 27 (2010).
\bibitem{weak-app3} A. Cho, Science {\bf 333}, 690 (2011).
\bibitem{weak-app4} H. Wiseman, Physics Letters A {\bf 311}, 285 (2003).
\bibitem{weak-app5} R. Mir, J. S. Lundeen, M. W. Mitchell, A. M. Steinberg, J. L. Garretson, and H. M. Wiseman, New Journal of Physics {\bf 9}, 287 (2007).
\bibitem{weak-app6} J. S. Lundeen and A. M. Steinberg, Phys. Rev. Lett. {\bf 102}, 020404 (2009).
\bibitem{weak-app7} A. M. Steinberg, Phys. Rev. Lett. {\bf 74}, 2405 (1995).
\bibitem{weak-app8} A. M. Steinberg, Phys. Rev. A {\bf 52}, 32 (1995).
\bibitem{weak-app9} Y.-S. Kim, J.-C. Lee, O. Kwon, and Y.-H. Kim, Nat. Phys. {\bf 8}, 117 (2012).
\bibitem{weak-app10} U. Singh, U. Mishra, and H. S. Dhar, Annals of Physics {\bf 350}, 50 (2014).
\bibitem{weak-app11} P. C. W. Davies, Phys. Rev. A {\bf 79}, 032103 (2009).
\bibitem{weak-app12} J. S. Lundeen, B. Sutherland, A. Patel, C. Stewart, and C. Bamber, Nature {\bf 474}, 188 (2011).
\bibitem{weak-app13} J. S. Lundeen and C. Bamber, Phys. Rev. Lett. {\bf 108}, 070402 (2012).
\bibitem{weak-app14} H. F. Hofmann, Phys. Rev. A {\bf 83}, 022106 (2011).
\bibitem{weak-rev} J. Dressel, M. Malik, F. M. Miatto, A. N. Jordan, and R. W. Boyd, Rev. Mod. Phys. {\bf 86}, 307 (2014).
\bibitem{dirac} P. A. M. Dirac, {\em The Principles of Quantum Mechanics} (Clarendon Press, 1981).\\ Dirac writes, {\em one might think one could measure a complex dynamical variable by measuring separately its real and pure imaginary parts. But this would involve two measurements or two observations, which would be alright in classical mechanics, but would not do in quantum mechanics, where two observations in general interfere with one another--it is not in general permissible to consider that two observations can be made exactly simultaneously, and if they are made in quick succession the first will usually disturb the state of the system and introduce an indeterminacy that will affect the second.}
\bibitem{nonhermitian-app1} H. Feshbach, Annals of Physics {\bf 5}, 357 (1958).
\bibitem{nonhermitian-app2} M. B. Plenio and P. L. Knight, Rev. Mod. Phys. {\bf 70}, 101 (1998).
\bibitem{nonhermitian-app3} Y. Aharonov, S. Massar, S. Popescu, J. Tollaksen, and L. Vaidman, Phys. Rev. Lett. {\bf 77}, 983 (1996).
\bibitem{nonhermitian-app4} P. A. M. Dirac, Rev. Mod. Phys. {\bf 17}, 195 (1945).
\bibitem{nonhermitian-app5} S. Chaturvedi, E. Ercolessi, G. Marmo, G. Morandi, N. Mukunda, and R. Simon, Journal of Physics A: Mathematical and General {\bf 39}, 1405 (2006).
\bibitem{nonhermitian-app6} L. M. Johansen, Phys. Rev. A {\bf 76}, 012119 (2007).
\bibitem{nonhermitian-app7} C. Bamber and J. S. Lundeen, Phys. Rev. Lett. {\bf 112}, 070405 (2014).
\bibitem{nonhermitian-app8} J. E. Moyal, Mathematical Proceedings of the Cambridge Philosophical Society {\bf 45}, 99 (1949).
\bibitem{nonhermitian-app9} A. Di Lorenzo, Phys. Rev. Lett. {\bf 110}, 010404 (2013).
\bibitem{nonhermitian-app10} A. Matzkin, Journal of Physics A: Mathematical and Theoretical {\bf 45}, 444023 (2012).
\bibitem{nonhermitian-meas} A. K. Pati, U. Singh, and U. Sinha, Phys. Rev. A {\bf 92}, 052120(2015).\\ The authors have proposed an experimentally verifiable method to measure the complex expectation value of an arbitrary non-hermitian operator. They showed that the average of a non-hermitian operator is a complex multiple of the weak value of the positive-semidefinite part of the non-hermitian operator.
\bibitem{bell-ineq} J. S. Bell, Physics 1, 195 (1964); J. F. Clauser, M. A. Horne, A. Shimony, and R. A. Holt, Phys.
Rev. Lett. {\bf 23}, 880 (1969); N. Brunner, D. Cavalcanti, S. Pironio, V. Scarani, and
S. Wehner, Rev. Mod. Phys. {\bf 86}, 419 (2014).
\bibitem{tsirelson}  B. S. Cirel'son, Lett. Math. Phys. {\bf 4}, 93 (1980).
\bibitem{ghz-argument} D. M. Greenberger, M. A. Horne, and A. Zeilinger, in {\em ``Bell's Theorem, Quantum Theory, and Conceptions of the Universe''}, edited by M. Kafatos (Kluwer, Dordrecht 1989).
\bibitem{watrous} John. Watrous, Theory of Quantum Information, Lecture 13. (2011)
\bibitem{note} $P_{\rho}=\frac{1}{2}[1/4+r_{x} (P_{+_{x}}-P_{-_{x}})+r_{y} (P_{+_{y}}-P_{-_{y}})+r_{z} (P_{+_{z}}-P_{-_{z}})]$ is normalized in the sense that $\sum_{\lambda=1}^8 P_{\rho}(\lambda) = 1$. However, one may argue that $P'_{\rho}=\frac{1}{8}[1+r_{x} (P_{+_{x}}-P_{-_{x}})+r_{y} (P_{+_{y}}-P_{-_{y}})+r_{z} (P_{+_{z}}-P_{-_{z}})]$ is also normalized since $\sum_{\lambda=1}^8 P'_{\rho}(\lambda) = 1$. But it should be true for arbitrary $\rho$. For $\rho=\ket{+_{x}}\bra{+_{x}}$ ($\overrightarrow{r}=(1,0,0)$), for example, while $\sum_{\lambda=1}^4 P_{\rho}(\lambda) = 1$, $\sum_{\lambda=1}^4 P'_{\rho}(\lambda) =\frac{5}{8} (<1)$.

\end{thebibliography}
\end{document}